\documentclass{article}

\usepackage{amsmath,amssymb,epsf,epsfig,a4wide}

\begin{document}

\begin{centering}

\vspace{0in}

{\Large {\bf Israel conditions for the Gauss-Bonnet theory and the Friedmann
equation on 
the brane universe}}

\vspace{0.4in}

{ \bf Elias Gravanis, \; Steven Willison} \\
\vspace{0.2in}
Department of Physics, King's College London,\\
Strand, London WC2R 2LS, United Kingdom.

\vspace{0.2in}
 {\bf Abstract}

\end{centering}

\vspace{0.2in}

{\small Assuming an Einstein-Gauss-Bonnet theory of gravitation in a ($D \geq
5$)-dimensional spacetime
with boundary, we consider the problem of the boundary dynamics given the matter
Lagrangian on it. The
resulting equation is applied in particular on the derivation of the Friedmann
equation of a 3-brane, 
understood as the non-orientable boundary of a 5d spacetime. We briefly discuss
the contradictory 
conclusions of the literature.}

\vspace{0.3in}

In the framework of Einstein gravitational theory, the cosmology of the
brane-world model
has been first studied in $\cite{Binetruy:1999ut}$. There the domain-wall/brane 
 is described as a 4d world-
volume slice of a 5d spacetime, which is of constant spatial curvature.  In
\cite{Mukohyama:1999wi}
it was shown that these solutions and the ones in $\cite{Kraus:1999it}$ 
 describe the same physical system which consists in a 3-surface 
of Robertson-Walker world-volume moving in the background of a 5d
AdS-Schwarzchild black hole
\footnote{More precisely, the construction is the following:
 the 3-surface is the boundary of ``half" such a space. Its location
with respect to the black hole is determined by the boundary dynamics.
One introduces a mirror image half, with respect to the
boundary, imposing also a ${\bf Z_2}$ symmetry to avoid overcounting. The
3-surface
becomes the non-orientable boundary of the new space. The evolution/motion of
the brane just
 recovers parts of the bulk spacetime which have
been cut off. This is because they are both of constant 3-curvature.}. Inverting 
this course one obtains a procedure 
for constructing brane cosmological solutions
that can be applied to
other gravitational theories.  Different
matter content on the 3-brane corresponds to a different 4d trajectory in the
bulk black hole spacetime.
The dynamics of the 4d trajectory is given by the appropriate Friedmann
equation, which
is obtained either by matching the bulk solutions around the brane or by viewing
the
3-wall as boundary of the spacetime(s) in the spirit of $\cite{Israel:rt}$, as
explained
nicely in $\cite{Chamblin:1999ya}$.

In order to extend these results in the case where 
the gravitational theory is corrected by  higher order terms in the curvature, 
it is at least convenient to
study the case where we add to the Einstein theory the Gauss-Bonnet term. This
particular combination of the Ricci scalar, tensor and Riemann tensor which
reads  $R^2-4 R_{\mu\nu}R^{\mu\nu}+R_{\mu\nu\rho\sigma}R^{\mu\nu\rho\sigma}$,
 leads to second order 
field equations which is not true for a general 
combination of these tensors. Its special properties in $D(\geq 5$) dimensions
are not unrelated to its nature as a topological invariant in 
4d $\cite{mathbook}$ or its origin from strings $\cite{Zwiebach:1985uq}$ 
\footnote{See also the discussion in $\cite{Myers:yn}$.}.

The black hole solution studied in $\cite{Boulware:1985wk}$ is the general
solution
of the Einstein-Gauss-Bonnet theory in 5d spacetime with constant spatial
curvature,
that respects Birkhoff's theorem $\cite{Charmousis:2002rc}$. We would like to
derive
the equation of motion of a 3-surface with constant spatial curvature, that
determines
its 4d trajectory in the 5d bulk or, in other words the Friedmann eq. of its
Robertson-Walker
world-volume, given the energy momentum tensor $T^{\mu}_{\nu}$ of the surface.

Put in general terms, the actual problem one is invited to solve is to find
the appropriate junction conditions around a surface filled with matter existing
in the spacetime, where a certain given gravitational theory determines its
metric.
As mentioned above, a way to do this is by treating each side of the surface as
boundary
of the respective part of the spacetime. Such a method is theoretically
appealing 
because  the surface truly separates two smooth manifolds. Also, the quantities
involved in the junction conditions are fundamental from that point of view,
thus naturally arising in the expressions. 

We study then the following problem: there is 
 a $D( \geq 5)$-spacetime M with a boundary $\partial M$.
In the interior of such a spacetime, $M-\partial M$, the ``bulk", 
we want the metric to be
determined by the field equations given by the 
minimization of the action of the Einstein-Gauss-Bonnet theory.
We assume that in the boundary there is localized a non-zero energy
momentum tensor. 
Now, in general, varying the action in the presence of the boundary, one is
left with normal derivatives of the variation of the
metric on the boundary which cannot go away by a partial integration.
One should then add a term in the action, local on the boundary, which
contributes similar terms that cancel the previous ones and
one is left with a well defined variational principle.
The needed boundary terms for the Einstein and Gauss-Bonnet 
theories were derived quite long ago and they belong
to the general class of Chern-Simons boundary terms (see references below).
Varying the full bulk plus boundary action with respect to the induced
metric on the boundary one obtains the equation of motion for it,
given in the text by equation
(\ref{thing}).
Note that no assumptions have been made and such an equation describes
the motion of a general boundary given the energy momentum tensor on it,
in a bulk which can contain matter as well. 
Combining then two manifolds like M and identifying the boundaries
one obtains the result for the surface mentioned at the beginning
of this paragraph, given in eq.(\ref{noz2}). This is the junction condition for an 
arbitrary surface in any bulk solution of the Einstein-Gauss-Bonnet gravity
in a $D(\geq 5)$-dimensional spacetime, which offers a more general
approach than previous work.  
If we impose a ${\mathbf Z_2}$ symmetry identifying
symmetrical points around the surface we obtain a still general result,
eq.(\ref{fin3}).
This equation looks like (\ref{thing}) but the assumed manifold is qualitatively
different;
we comment on that in the text.
As an application of the junction conditions we treat the
specific problem mentioned above, deriving the Friedmann equation/equation
of motion of a ``brane universe" assuming Einstein-Gauss-Bonnet
gravity in the bulk, a problem already solved in $\cite{Charmousis:2002rc}$,
resulting in 
(\ref{expl}).

Assume then that the gravity is described by Einstein-Gauss-Bonnet theory in the
bulk of
a ($D \geq 5$)-dimensional spacetime which 
 is a manifold $M$ with boundary $\partial M$. The theory
takes an elegant form and calculations are easier if we write everything in
terms
of differential forms. We use the notation of $\cite{Myers:yn}$.
Let $E^A$ be the normalized basis of 1-forms in terms of which the metric is 
$g= \eta_{AB} E^A \otimes E^B$
with $\eta_{AB}=(-+++..)$. Also define
\begin{eqnarray}
e_{A_1..A_m}=\frac{1}{(D-m)!} \, \epsilon_{A_1..A_m A_{m+1}..A_D} E^{A_{m+1}}
\wedge .. \wedge E^{A_D}
\end{eqnarray}
 where
$\epsilon_{A_1..A_D}$ is the completely antisymmetric tensor with the
normalization $\epsilon_{0..D-1}=1$.
The curvature 2-form $\Omega^{AB}$ in terms of the connection $\omega^{AB}$ and 
the Riemann tensor $R^A_{\; BCD} $ is
$\Omega^A_{\;B}=d\omega^A_{\;B}+\omega^A_{\;C} \wedge \omega^C_{\;B}=\frac{1}{2}
R^A_{\; BCD} E^C \wedge E^D$.
The second fundamental form $\theta^{AB}$ is defined as follows:
if we introduce  Gaussian normal coordinates $(x,w)$ chosen so that the
(time-like) boundary 
$\partial M$ is given by $w=0$, the
metric is written as
$ds^2=dw^2+ \gamma_{\mu\nu}(x,w) dx^{\mu} dx^{\nu}$.
We define $\omega_0$ as the connection of the product metric that agrees
 with the previous metric at $\partial M$: $ds^2= dw^2+\gamma_{\mu\nu}(x,0)
dx^{\mu} dx^{\nu}$.
Clearly this connection has non-zero components only tangentially on the
boundary. Then
\begin{eqnarray}
\theta= \omega- \omega_0
\end{eqnarray}
or in terms of the extrinsic curvature $K^{AB}$ 
\begin{eqnarray}
\theta^{AB}=\theta^{AB}_{\quad C} E^C=(N^A K^B_C- N^B K^A_C) E^C= 2 N^{[A}
K^{B]}_C E^C
\end{eqnarray}
$N^A$ is the normal vector on the boundary.
Explicitly the extrinsic curvature reads $K^A_B=-(\eta^{AC}-N^AN^C) \nabla_C
N_B$ or in
$(x,w)$ coordinates
$K_{\mu\nu}=-\frac{1}{2} \; \partial \gamma_{\mu\nu}/ \partial w$.
Note that $\theta^{AB}$ has only mixed (normal - tangential on the boundary)
components
due to its antisymmetry and the property  $N^A K_{AB}=N^B K_{AB}=0$.

Then the action, discussed explicitly in $\cite{Myers:yn}$
(see also $\cite{Eguchi:jx}$) 
, that contains appropriate
boundary terms so that the normal derivatives of the variations of the metric
cancel identically, can be written
\begin{eqnarray}  \label{action}
S=\int_M -2\Lambda \; e+\Omega^{AB} \wedge e_{AB} + \alpha \; \Omega^{AB} \wedge
\Omega^{CD} \wedge e_{ABCD}- \\ \nonumber
- \int_{\partial M} \theta^{AB} \wedge e_{AB}+ \alpha \; 2 \theta^{AB} \wedge 
( \Omega^{CD} - \frac{2}{3} \theta^C_E \wedge \theta^{ED}) \wedge e_{ABCD}
\end{eqnarray}
where we have also introduced a bulk cosmological constant $\Lambda$. In
the bulk part of the action the linear term
in the curvature 2-form is the Einstein-Hilbert action and the quadratic is the
Gauss-Bonnet term.
We designate $\alpha$ the coupling of the Gauss-Bonnet term which has units
(length)$^2$.
If the higher curvature term is thought of as originating from string theory
$\alpha$
is proportional to the Regge slope $\alpha'$. The boundary part is the
Chern-Simons
term mentioned in the introductory part of our paper.
\footnote{We are mainly interested in Chern class type actions in view of their
desirable
properties. For a construction of the boundary term for a bulk Lagrangian formed
as
a general polynomial of the Riemann tensor
see $\cite{Barvinsky:1995dp}$. }

Varying with respect to the basis forms while keeping the connection fixed we
have
\begin{eqnarray} \label{var}
\delta_E S=\int_M \delta E^F \wedge (-2 \Lambda \; e_F + \Omega^{AB} \wedge
e_{ABF} + \alpha \;
\Omega^{AB} \wedge \Omega^{CD} \wedge e_{ABCDF})+ \\ 
+ \int_{\partial M} \delta E^F \wedge (\theta^{AB} \wedge e_{ABF}+\alpha \; 2
\theta^{AB} \wedge 
( \Omega^{CD} - \frac{2}{3} (\theta \wedge \theta)^{CD}) \wedge e_{ABCDF})
\nonumber
\end{eqnarray}
If there is no matter in the bulk, the bulk volume integral vanishes giving the
field equation outside the
boundary. Assuming that there is matter on the boundary, the boundary integral
above equals the energy momentum
tensor of the matter on it coming from the variation of the matter Lagrangian. 

In order to write this in an explicit form in terms of the extrinsic curvature
and the intrinsic curvature
tensors of the boundary, we first note 
that by $\omega=\omega_0+\theta$ we have
\begin{equation}
\Omega^{AB}=\Omega_0^{AB}+d\theta^{AB}+(\omega_0 \wedge \theta)^{AB}+
(\theta \wedge \omega_0)^{AB} +(\theta \wedge \theta)^{AB}
\end{equation}
where $\Omega_0=d\omega_0+ \omega_0 \wedge \omega_0$ is the intrinsic curvature
2-form of the boundary.
Using that in (\ref{var}) we see that the boundary terms can be written as
\begin{eqnarray} \label{var2}
\int_{\partial M} \delta E^F \wedge (\theta^{AB} \wedge e_{ABF}+\alpha \; 2
\theta^{AB} \wedge 
( \Omega_0^{CD} + \frac{1}{3} (\theta \wedge \theta)^{CD}) \wedge e_{ABCDF}) 
\end{eqnarray} 
as the other terms have to have an index on the normal direction that
contributes zero
as there is a factor $\theta^{AB}$ already in the expression.
Using the second of the identities
\begin{eqnarray}
E^C \wedge e_{ABF}= \frac{3!}{2!} \delta^C_{[F} e_{AB]}, \quad 
E^G \wedge E^H \wedge E^I \wedge e_{ABCDF}= \frac{5!}{2!} \delta^I_{[F}
\delta^H_D \delta^G_C e_{AB]}=
\frac{5!}{2!} \delta^G_{[A} \delta^H_B \delta^I_C e_{DF]},
\end{eqnarray}
 the first Gauss-Bonnet boundary term gives
\begin{eqnarray}
2 \alpha \, \theta^{AB} \wedge \Omega_0^{CD} \wedge e_{ABCDF}= 
\alpha \, \theta^{AB}_{\; G} \; R^{CD}_{0 \quad HI} \,
\frac{5!}{2!} \; \delta^G_{[A} \delta^H_B \delta^I_C e_{DF]}=
\alpha \, N^A 5! K^B_{[A} \; R^{CD}_{0 \quad BC} e_{DF]}
\end{eqnarray}
Given that the first three indices are orthogonal to $N^A$ we can write
\footnote{Our convention is $R^{\mu}_{\; \nu\rho\sigma}= \partial_{\rho}
\Gamma^{\mu}_{\nu\sigma}+..$ .}
\begin{eqnarray} 
&& \alpha \, N^A5! K^B_{[A} \; R^{CD}_{0 \quad BC} e_{DF]}=\alpha \, N^A \,
3!\,2 ( K^B_{[B}R^{CD}_{0 \quad CD]} e_{FA}-
K^B_{[F}R^{CD}_{0 \quad BC]} e_{DA} +  \\ && +K^B_{[D}R^{CD}_{0 \quad FB]}
e_{CA}-
K^B_{[C}R^{CD}_{0 \quad DF]} e_{BA} ) \nonumber
\end{eqnarray}
which is easy to see performing the contractions that it takes the form
\begin{eqnarray}
-4\alpha \, N^A \, e_{AB} ([4 (KR_0)^B_F+2K^D_C \, R^{CB}_{0 \quad DF}-2K R^B_{0
\,F }- K^B_F R_0 ]+
\delta^B_F [K R_0 -2 Tr(K R_0)])
\end{eqnarray}
Doing the same for the $\theta \wedge \theta$ term, 
using  $(\theta \wedge \theta)^{CD}=- K^C_H K^D_I \, E^H \wedge E^I$, 
we can finally write the integrand of (\ref{var2}) in the form
\begin{eqnarray} \label{thing} 
&&  \delta E^F \wedge N^A e_{AB} \; [2 (K^B_F- \delta^B_F K) 
+4 \, \alpha ( Q^B_F- \frac{1}{3} \, \delta^B_F Q^C_C )], \\
&&Q^B_F= -2 K^D_C R^{CB}_{0\quad DF}-4 (R_0 K)^B_F +2 K R^B_{0F}+ R_0 K^B_F+
\nonumber \\ \nonumber && \qquad+
 K^B_F (TrK^2-K^2)+
2K (K^2)^B_F -2 (K^3)^B_F \nonumber
\end{eqnarray}
The dynamics of the each boundary is described by setting the quantity in square
brackets equal to 
$-2 T^B_F$, for $N^A$ oriented outwards the space
 according to Stokes theorem and
 $T^B_F$ is boundary matter energy momentum tensor. The 
normalization is fixed by the 
the bulk part of the (\ref{var}) which reads $-2( \delta^B_F \Lambda +
G^B_F+..)$, where
$G^B_F$ is the Einstein tensor and we use a convention such that the field
equations have
the form $G^B_F+..=T^B_F$.

With this result at hand we can now treat the problem of a surface in the
spacetime in such a theory.
We can think of the spacetime as two manifolds $M_i,i=1,2$ whose boundaries are
identified. The
normal vector of the common boundary with respect to each $M_i,i=1,2$ are equal
up to a sign, $N_1^A=-
N_2^A$. The extrinsic curvatures,
describing
the surface in different spacetimes, will in general differ. 
The variation of the gravity action with
respect to the induced metric 
is going to give contributions from both sides of the common boundary
leading to the
following equations of motion for the surface
\begin{eqnarray} \label{noz2}
 (K^B_F- \delta^B_F K)_1 
+2 \, \alpha ( Q^B_F- \frac{1}{3} \, \delta^B_F Q^C_C )_1+
 (K^B_F- \delta^B_F K)_2 
+2 \, \alpha ( Q^B_F- \frac{1}{3} \, \delta^B_F Q^C_C )_2=T^B_{F}
\end{eqnarray}
The suffixes denote that the quantities should be calculated in the 
indicated side. Note that the intrinsic in the boundary curvature tensors
in the definition of $Q^A_B$ do not depend on the side. Note that
as it is natural the normal vectors are taken to be oriented 
inwards the respective $M_i$.

A particularly interesting restriction to the above is
when we identify points of $M_1$ with points of $M_2$ around the surface.
At each point in the surface, taken to be the
origin of the normal direction $w$ we identify symmetrical points.
 The normal vector at the
origin has to have both directions and the origin can be understood as
a non-orientable boundary of the spacetime. Practically
 one, integrating over the boundary, should 
integrate over both directions. The point is that
due to the $\mathbf Z_2$ symmetry, both the normal vector and the extrinsic
curvature change sign
going from one ``side" of the boundary to the other, leading to a total factor
of 2 in the final 
result. 
Also, as it is natural to take the normal vector oriented inwards each
half-space,
there is an additional minus sign.
Then the equation of boundary motion or junction condition under the $\mathbf
Z_2$ symmetry
restriction takes the form
\begin{eqnarray} \label{fin3}
2 (K^B_F- \delta^B_F K) 
+4 \, \alpha ( Q^B_F- \frac{1}{3} \, \delta^B_F Q^C_C )=T^B_{F}
\end{eqnarray}

We have considered three cases: we can simply have a spacetime with a free
boundary;
the common boundary described above; and the specific case of 
a non-orientable boundary with ${\bf Z_2}$ symmetry.
The qualitative difference between these three case
leads to a distinction between the general kind of theories to which they can be
applied. In the
first case one demands, roughly, that the generic metric function is defined
on the half-line, $g(w), w \geq 0$, in the third that its continuation
beyond the origin is given by $g(-w)=g(w)$, while in the second 
no particular restriction is imposed. With ${\bf Z_2}$ symmetry, the second
derivative in any expression 
is going to produce a delta function at the origin where the non-orientable
boundary is located. This means that there only first derivatives should arise
on the boundary terms of the theory which describes the gravitation in the bulk,
that is the theory should have second order field equations. As noted earlier,
Gauss-Bonnet is the only combination with this property,
in the kind of theories
which are built by the invariant squares of the curvature
tensors, so it is the only one that applies under
the ${\bf Z_2}$ symmetry. In the other cases a more general combination,
whose boundary term can be constructed, should apply.

Let us now apply these results to the case of a 3-wall of constant spatial
curvature in
the background of the particular black hole solution of the
Einstein-Gauss-Bonnet 5d spacetime
of constant spatial curvature
mentioned earlier. 
Its line element can be written in the form
\begin{eqnarray} \label{bh}
&&ds^2=-f(y) \; dt^2+ \frac{dy^2}{f(y)}+y^2 dx^2, \\ \nonumber
&&dx^2=\frac{dr^2}{1-kr^2}+r^2(d\theta^2+\sin^2\theta d\phi^2) \\ \nonumber
&&f(y)=k+\frac{ y^2}{4\alpha} \left( 1 \pm \sqrt{1+  
\frac{4\alpha \Lambda}{3}+\frac{8 \alpha \mu}{ y^4}} \right)  
\end{eqnarray}
where $\mu$ is the gravitational mass of the black hole.
$\Lambda$ is positive for de Sitter
space and should be bounded as $\Lambda \geq - 3/4 \alpha$.
For the sake of the result as well as for its usefulness,  
we use the method of \cite{Mukohyama:1999wi} to transform from the black hole
coordinates to
the ones built around the trajectory of the wall, where the coordinate $w$ is
the proper-``time" of the
spacelike geodesics that cross the trajectory vertically.
Then one finds that the metric (\ref{bh}) can be also written
in another familiar form that reads
\begin{eqnarray} \label{wall}
ds^2=-\frac{\psi^2(\tau,w)}{ \varphi (\tau ,w)} d\tau^2+ a^2(\tau) \varphi
(\tau,w) dx^2+dw^2
\end{eqnarray} 
where the function $\varphi$ is given implicitly by the equation
\begin{eqnarray} \label{phi}
\int^{\varphi}_1 \frac{dx}{V^{1/2}}= \pm 2w
\end{eqnarray}
where $V$ is given by
\begin{eqnarray}
V=x\left(H^2+ \frac{f(a\sqrt{x})}{a^2} \right)=x \left( H^2+ \frac{k}{a^2}
\right)+ \frac{x^2}{4 \alpha} \;
\left( 1 \pm \sqrt{1+\frac{4 \alpha \Lambda}{3} +\frac{8 \alpha \mu}{ a^4 x^2}}
\right) 
\end{eqnarray}
The Hubble parameter is $H=\dot a(\tau) / a(\tau)$ and
$\psi(\tau,w)=\varphi(\tau,w)+\frac{1}{2H} ( \frac{\partial \varphi}{\partial
\tau} )_w$,
where the derivative is given implicitly by (\ref{phi}) and
\begin{equation}
\frac{1}{2H} ( \frac{\partial \varphi}{\partial \tau})_w = \frac{
V^{1/2}|_{x=\varphi } }{2}
\int^{\varphi}_1   \frac{dx}{V^{3/2}}   \left( x \left( \dot H-\frac{k}{a^2}
\right) 
\mp \frac{2\mu}{a^4}  \left(  1+\frac{4\alpha \Lambda}{3} +\frac{8 \alpha \mu}{
a^4 x^2} \right)^{-3/2} \right)
\end{equation}
Note that $\varphi (\tau,0)=1$ and $ \partial_{\tau} \varphi (\tau,0)=0$ so
$\psi(\tau,0)=1$. That is
the induced metric at $w=0$ is Robertson-Walker with scale factor $a(\tau)$.

We define $S^2(\tau,w)=a^2(\tau) \varphi (\tau,w)$
and $N^2(\tau,w)= \psi^2(\tau,w)/  \varphi (\tau ,w)$ and calculate the
00-component of 
(\ref{fin3}) to obtain
\begin{eqnarray} \label{expl}
2\frac{S'_0}{S_0} \left( 3 +12 \alpha \left( \frac{\dot S_0^2}{N_0^2
S_0^2}+\frac{k}{S_0^2} \right)
-4 \alpha \frac{{S'}^2_0}{S_0^2} \right)=  -\rho
\end{eqnarray}
where $S'_0=S'(0^+)=-S'(0^-)$ fixed at that value from the product of the normal
vector with the
extrinsic curvature as explained earlier. Note that $S_0=S(\tau,0)=a(\tau)$ and 
$H=\dot S_0/ S_0= \dot a(\tau)/a(\tau)$
with $N_0=N(\tau,0)=1$.

Taking the square of that equation, we see that we only need the quantity
${S'}^2_0/S_0^2$, which is obtained
by using equation (\ref{phi}) and the definition of $S(\tau,w)$. We have
\begin{eqnarray} \label{basicdis}
&&\frac{{S'_0}^2}{S_0^2}=\frac{1}{4} ( \frac{\partial \varphi} { \partial w}
)_{w=0}^2=V(x=1)=
H^2+ \frac{f(a)}{a^2}= \\ && \quad =H^2+ \frac{k}{a^2} + \frac{1}{4 \alpha} \;  
\left( 1 \pm \sqrt{1+\frac{4 \alpha \Lambda}{3} +\frac{8 \alpha \mu}{ a^4 }}
\right) \nonumber
\end{eqnarray}
Substituting in (\ref{expl}) we obtain
 \begin{eqnarray} \label{h}
&&4 (H^2+\frac{k}{a^2}- {\phi}) (3+ 8 \alpha  
(H^2+\frac{k}{a^2}) + 4 \alpha {\phi})^2 = {\rho}^2, \\
&& \phi= - \frac{1}{4 \alpha} \;  
\left( 1 \pm \sqrt{1+\frac{4 \alpha \Lambda}{3}
+\frac{8 \alpha \mu}{ a^4 }}\right) \nonumber
\end{eqnarray}

The single real solution of the equation above 
with a smooth limit $\alpha \to 0$
(corresponding to
the minus sign choice for the functions $f(y)$ and $\phi$\,),
can be written in the form
\begin{eqnarray} \label{sol}
H^2= -\frac{1}{4 \alpha} + \frac{(\phi+ \frac{1}{4 \alpha})^2}{Q^{2/3}}+
\frac{Q^{2/3}}{4}-\frac{k}{a^2}, \\
Q=\frac{1}{4 \alpha} \left( \rho+ \sqrt{\rho^2+ 128 \alpha^2 (\phi+ 
\frac{1}{4 \alpha} )^3} \right) \nonumber
\end{eqnarray}
The first order corrections to the result of $\cite{Binetruy:1999ut}$ can be
read from
\begin{eqnarray}
H^2+\frac{k}{a^2}=\left(1-\frac{4\alpha\Lambda}{3}-\frac{72\alpha\mu}{a^4}
\right) \frac{\rho^2}{36}
-\alpha \frac{\rho^4}{243}
+\frac{\Lambda}{6} \left(1-\frac{\alpha\Lambda}{3} \right) 
+ \frac{\mu(1-6\alpha\Lambda)}{a^4}-\alpha \frac{2\mu^2}{a^8}
\end{eqnarray}
Shifting the energy density, $\rho=\eta+\varrho\,$,  and tuning $\eta$ so that
to have 
a vanishing 4d cosmological constant, we obtain
\begin{equation}
H^2+\frac{k}{a^2}=\frac{\sqrt{-\Lambda}}{3 \sqrt{6}} 
\left(1+  \frac{\alpha\Lambda}{2}-\frac{8\alpha \mu}{a^4} \right) \varrho+
\frac{\mu(1+\frac{2}{3} \alpha \Lambda)}{a^4}-\frac{2\alpha\mu^2}{a^8} 
+ {\cal O}(\varrho^2)
\end{equation}
The effective 4d coupling constant has become scale factor dependent.
\footnote{For the plus sign choice in the definition of $f(y)$ the solution
(\ref{bh})
is classically unstable $\cite{Boulware:1985wk}$. A simple analysis of the
equation (\ref{h})
shows that in this case (where $\phi+ \frac{1}{4\alpha}<0$), the condition for a
single real
solution is $\rho^2+ 128 \alpha^2 (\phi+ 
\frac{1}{4 \alpha} )^3>0$. Otherwise there are three real solutions.
 For $\phi+\frac{1}{4\alpha}>0$
there is only one real solution, eq. (\ref{sol}), without constraint.}

Now let us see how the same result arises when we treat the 3-wall as a body in
the bulk. Calculating the
00-component of the bulk field equations we obtain
\begin{equation}
\frac{3S''}{S}-
12 \alpha \frac{ S'' {S'}^2}{S^3}+ 12 \alpha  \frac{k S''}{S^3}+
12 \alpha \frac{ S'' \dot S^2}{N^2S^3}+..=- \rho \delta(w)
\end{equation}
where dots contain terms involving only first 
$w$-derivatives and $S'=\partial_w S$ and $\dot S=\partial_{\tau} S$. 

We integrate both sides
in an infinitesimal region around zero. The first derivative of the metric
functions, such as
$\varphi$, are taken to change sign passing through the point $w=0$ as is
actually implied by
(\ref{phi}). With that in mind one can write
\begin{equation}
\partial_w \left(\frac{3S'}{S}- 4 \alpha \frac{  {S'}^3}{S^3}
 +   12 \alpha \frac{ k S'}{S^3}
+ 12 \alpha \frac{S' \dot S^2}{N^2S^3} \right)+..=- \rho \delta(w)
\end{equation}
where dots are first derivative terms that contribute zero to the integral. Then
\begin{equation} \label{dis}
\frac{3[S_0']}{S_0}- 4 \alpha \frac{  [{S'}_0^3]}{S_0^3}+
 12 \alpha  \frac{ k [S_0']}{S^3_0}+ 12 \alpha \frac{ [S_0'] \dot
S_0^2}{N_0^2S_0^3}=-\rho
\end{equation}
where  $[S']=S'(0^+)-S'(0^-)$. $\bf Z_2$ symmetry 
implies that $S'(0^+)=-S'(0^-)$ 
that is $[S']=2S'(0^+)=2S'_0$ and
${S'}^2(0^+)={S'}^2(0^-)={S'}^2_0$. Then, we obtain the same Friedmann equation.
This result is
agreement with the results of $\cite{Charmousis:2002rc}$ 
and $\cite{Davis:2002gn}$ and the analysis of $\cite{Low:2000pq}$.

In $\cite{Deruelle:2000ge}$ it was argued that the quantities that appear in
these formulas
make the expression not well defined from the point of view of distributions.
This due to 
the existence of a term involving ${S'}^2 S''$, as ${S'}^2$ is a discontinuous
function. On the
other hand, this can be combined to $({S'}^3)'$ which is well defined and
still a delta function, as the derivative of both the sign function and its cube
behave as
delta functions multiplied with smooth functions. A real difficulty would arise
only 
in the case of product of distributions, as in the loop calculations in quantum
field theory,
where this leads to an introduction of a cutoff. This is the conclusion of
$\cite{Deruelle:2000ge}$
for the problem of the Friedmann equation on the brane in the Gauss-Bonnet
theory, where
based on that, it is argued that the equation does not change, only the coupling
constants, as
in the renormalization of loop graphs.

 The same is suggested in $\cite{Kim:2000pz}$, where
finite results have been found, by treating the sign function squared
$\epsilon^2(w)$, in products
with the delta function, as a constant function at the value one. On the other
hand 
integrating the functions $\epsilon(w)$ and $\epsilon^2(w)$ with the delta
function, it is clear that
the usual rules are not obeyed when one assigns prescribed values to them at
$w=0$, through
irrelevant limiting procedures. 

Concluding the main discussion, we would like to emphasize that the procedure of
obtaining the junction conditions
by the appropriate boundary term makes clear that the domain wall in an
Einstein-Gauss-Bonnet theory
is a surface of zero thickness as much as it is in the Einstein theory. This is
because 
one is working directly with the surface of the boundary and there is no room
for any regularization in the normal direction.

Finally, we briefly discuss theories with additional bulk fields.
 Assume for example that the gravity
includes the dilaton field and its action is given by 
\begin{eqnarray}
S=\int d^5x \left( -2\Lambda-\frac{1}{2} (\nabla \Phi)^2
+R+ \alpha\, h(\Phi) (R^2-4
R_{\mu\nu}R^{\mu\nu}+R_{\mu\nu\rho\sigma}R^{\mu\nu\rho\sigma})
 \right)
\end{eqnarray}
This is just one convenient choice so that to obtain a Friedmann equation with a
dilaton type
field present. A general solution for the cosmology on the brane in this theory,
is
equivalent to this 3-surface of constant spatial curvature, moving
in a general background in the 5d spacetime with the same property.
Assume then, that a certain general static solution of constant spatial
curvature, as in (\ref{bh}) with a different $f(y)$, is known, as well as the
form 
of the dilaton $\Phi=\Phi(y)$
\footnote{To our knowledge there are no analytically  known solutions of this
kind. 
It will be harder to find cosmological solutions of such a theory 
using a time dependent metric, e.g. in Gaussian normal coordinates
along the lines of $\cite{Binetruy:1999ut}$.
We derive the Friedmann equation in this case out of theoretical interest. It
can of course be applied to numerical solutions. Solutions of the
type we are interested in have been studied for the case of 4d in
$\cite{Kanti:1995vq}$.}.
 By integrating the 00-component of the field equations,
calculated
for the line element in the Gaussian normal coordinates form, 
around $w=0$ and following steps explained above, we obtain
\begin{equation}
4 \left( H^2+\frac{f(a)}{a^2} \right)
  \left( -3+ 4 \alpha \left(h+ 3 a \frac{dh}{da} \right)
  \left( 
 H^2+\frac{f(a)}{a^2} -3 \left(  H^2+\frac{k}{a^2} \right)  \right) 
+ 24 \alpha a \frac{dh}{da} \frac{k}{a^2} \right)^2=\rho^2
\end{equation}
where $a=a(\tau)$ as above and $h=h(\Phi(a))$ and we have used the relation
${S'}_0^2/S_0^2=H^2+f(a)/a^2$ as
shown above (eq. (\ref{basicdis}). For $h=1$ this goes over to equation
(\ref{h}).
It is still a cubic equation with respect to the Hubble parameter $H^2$ but with 
scale factor dependent coefficients.

For the case of a bulk form-field, practically only the black hole changes so
that the Hubble parameter
still satisfies a cubic equation similar to $(\ref{h})$. 
Such cases has been studied
at least in $\cite{Lidsey:2002zw}$.

$\mathbf {Note \; added}$: While this work had been at the final stages S. C.
Davis
reported similar analysis in $\cite{Davis:2002gn}$. 

$\mathbf{ \qquad Acknowledgements}$

The work of EG was supported by a King's College Research Scholarship (KRS). The
work of SW was supported by EPSRC. This work is also partially supported
by the E.U. (contract ref. HPRN-CT-2000-00152). We thank Nick Mavromatos (King's
College London)
and John Rizos (Ioannina Univ.) for useful
discussions.
We also thank 
E. Papantonopoulos (NTU Athens) for
discussions at the early stages of this work.

\end{document}